\documentclass[9pt]{extarticle}
\usepackage{spconf} 

\usepackage{multicol}

\newcommand{\mb}{\mathbf}
\newcommand{\mc}{\mathcal}

\refstepcounter{equation} 
\usepackage{amsfonts,upgreek,pifont,amssymb,enumerate,color,algorithm,theorem,algorithmicx}

\usepackage{algpseudocode}
\usepackage{cite,graphicx,subfigure,epstopdf}

\usepackage{mathtools}
\usepackage[nameinlink]{cleveref}

\newtheorem{Lemma}{Lemma}
\newtheorem{Prop}{Proposition}

\begin{document}
\title{Latency-Minimized Design of Secure Transmissions in UAV-Aided Communications}
\name{Xiongwei Wu$^{1,2}$, Qiang Li$^{3}$, Yawei Lu$^{2,4}$, H. Vincent Poor$^{2}$, Victor C. M. Leung$^{5,6}$ and P. C. Ching$^1$}
\address{$^1$Department of Electronic Engineering, The Chinese University of Hong Kong, Hong Kong SAR, China\\
$^{2}$ Department of Electrical Engineering, Princeton University, Princeton, USA\\
$^{3}$School of Info. $\&$ Comm. Eng., University of Electronic Science and Technology of China, Chengdu, China\\
$^{4}$ Department of Electronic Engineering, Tsinghua University, Beijing, China\\
$^5$College of Computer Science \& Software Engineering, Shenzhen University, Shenzhen, China\\
$^6$Department of Electrical \& Computer Engineering, The University of British Columbia, Canada
\thanks{This work was supported in part by the Global Scholarship Programme for Research Excellence from CUHK, and in part by the U.S. National Science Foundation under Grants CCF-0939370 and CCF-1513915.}
}
\maketitle
\begin{abstract}
  Unmanned aerial vehicles (UAVs) can be utilized as aerial base stations to provide communication service  for remote mobile users due to their high mobility and flexible deployment. However, the line-of-sight (LoS) wireless links are vulnerable to be intercepted by the eavesdropper (Eve), which presents a major challenge for UAV-aided communications. In this paper, we propose a latency-minimized transmission scheme for satisfying legitimate users' (LUs') content requests securely against Eve. By leveraging physical-layer security (PLS) techniques, we formulate a transmission latency minimization problem by jointly optimizing the UAV trajectory and user association. The resulting problem is a mixed-integer nonlinear program (MINLP), which is known to be NP hard. Furthermore, the dimension of optimization variables is indeterminate, which again makes our problem very challenging. To efficiently address this, we utilize bisection to search for the minimum transmission delay and introduce a variational penalty method to address the associated subproblem via an inexact block coordinate descent approach. Moreover, we present a characterization for the optimal solution. Simulation results are provided to demonstrate the superior performance of the proposed design. 
\end{abstract}
\begin{keywords}
Unmanned Aerial Vehicles, Physical-layer Security, Penalty Method, Latency
\end{keywords}
\section{Introduction}
Unmanned aerial vehicle (UAV)-aided communication has been widely envisioned as a critical infrastructure for future Internet of Things (IoT) networks \cite{chen2019artificial,challita2019machine}. Different from conventional ground base stations (BSs), UAVs are capable of flexible movement, on-demand deployment as well as high probability of line-of-sight (LoS) wireless links \cite{zeng2016wireless}. Thus, it has many appealing applications, e.g., serving as relays to provide ubiquitous communications for remote mobile users, resuming Internet access for special areas, aggregating or delivering content items in IoT applications 
\cite{chen2017caching,zeng2016wireless,qian2019user}.

However, due to the broadcast nature of line-of-slight (LoS) wireless channels between UAVs and legitimate users (LUs), content delivery is vulnerable to information leakage and likely to be wiretapped by eavesdropper (Eve) \cite{hua2019energy}. In particular, UAVs may move around and get close to Eve when supporting content transmissions for LUs. This fact gives a new challenge to develop secure transmission designs for satisfying users' demands. Some preliminary works have been devoted to secure UAV-aided communications by leveraging physical-layer security (PLS). Specifically, 
the research in \cite{cai2018dual} focused on improving the worst-case secrecy rate among mobile users by generating a jamming signal in order to confound Eves. The work in \cite{hua2019energy} studied secrecy energy efficiency in multiple UAV-aided wireless networks by leveraging cooperation. Moreover, covert communications for UAV-aided networks was examined in \cite{zhouOptimization2019joint}, where the averaged covert transmission rate was maximized by optimizing trajectory and transmission power. Unfortunately, transmission latency for UAV-aided communications, which is regarded as one of core concerns in future IoT networks \cite{chen2019artificial}, has not been adequately studied so far. 
Only very few studies have paid attention to this issue. Prior works in \cite{gong2018flight,zeng2018trajectory} proposed transmission designs to minimize latency of either data collection or content downloads. 
None of the above latency-minimized studies takes into account transmission security. Thus, secure transmission design with a latency-minimized objective remains an open issue. 

To bridge the research gap identified above, in this paper, we develop a latency-minimized design for LUs while ensuring all requested content is securely transmitted in the presence of an Eve. Specifically, the UAV serves as an aerial base station to provide service for LUs, which may be overheard by Eve. Taking advantage of UAV movement and user association, we exploit the PLS technique to generate degraded channels for Eve and thereby 
avoid information leakage. Thus, a transmission latency minimization problem is formulated by jointly optimizing the UAV trajectory and user association strategy. 
The resulting problem is a mixed integer nonlinear program, where the dimension of variables also needs to be optimized.  
To deal with such a complicated problem, we introduce a bisection approach to efficiently find an optimized transmission latency by addressing its corresponding subproblem with fixed transmission latency in each iteration. 
Notably, instead of relaxing the binary variables into continuous ones like most extant studies, a variational penalty approach is proposed. Simulation results demonstrate significant advantages over the continuous relaxation method. Moreover, the proposed design can effectively find a low-latency transmission design through comparisons with baselines.







\section{System Model}
\begin{figure}[h]
  \centering
  \includegraphics[scale=0.4]{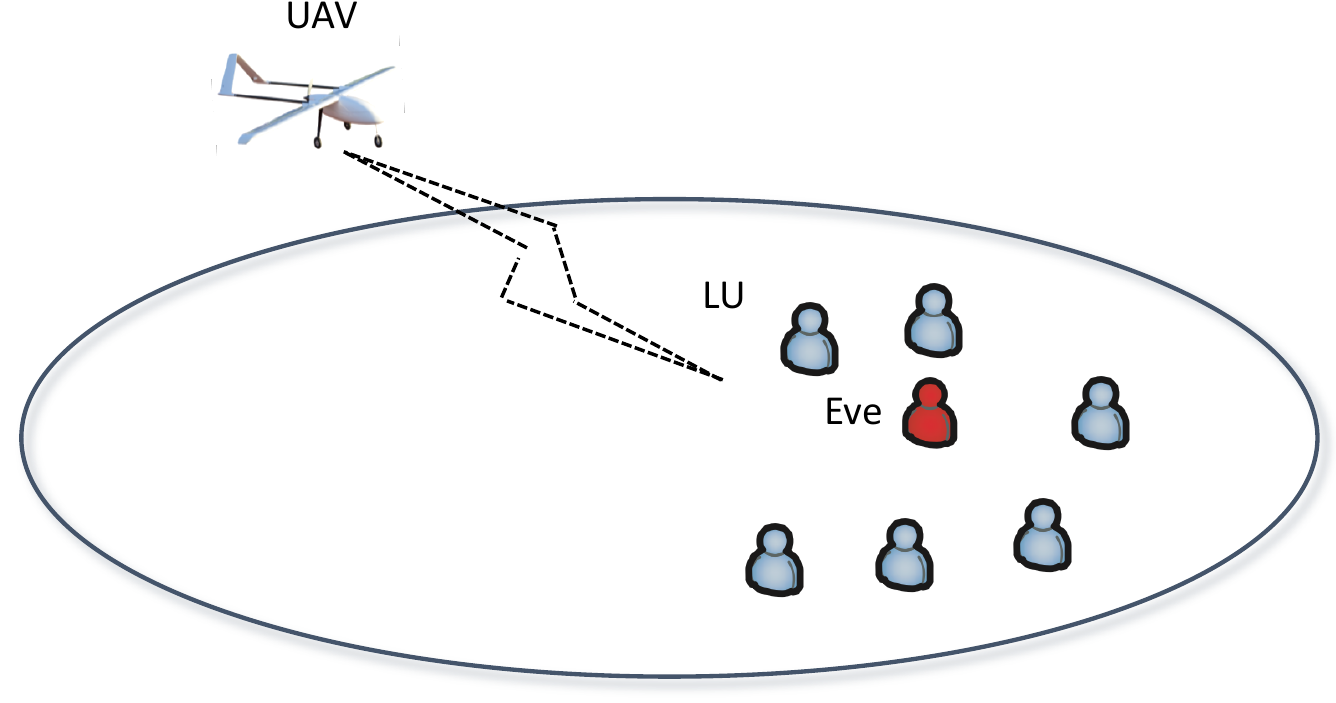}
  \caption{An illustrative model for UAV-aided communications}
  \label{system}
\end{figure}

As shown in Fig. \ref{system}, we consider a UAV-aided secure communications, where the UAV provides communication service for $K$ LUs. Specifically, each LU can request content items from the UAV by following certain content preference distribution (e.g., Zipf distribution in \cite{li2018hierarchical}).
Note that Eve is located around LUs, which can wiretap information delivery between the UAV and LUs.  Let $\mc K = \{1,2,\cdots, K\}$ be the set of indices of LUs.  
For ease of discussion, each LU requires only one content each time; and all contents are considered to be equal in size, each of which contains $s$ bits.
After users' requests are revealed, the UAV is required to securely ferry those contents to users with satisfactory quality of service. In this paper, we focus on designing efficient secure transmission scheme to minimize transmission delay. 

Consider that UAV periodically flies over the area of interest with period of $T$ \cite{hua2019energy}.
For ease of discussion, the continuous time period is equally partitioned into discrete time slots $\mc N = \{ 1, 2, \cdots, N\}$, with $N = T/\tau$ and $\tau$ being the discrete time interval. In general, $\tau$ needs to be sufficiently small so as to ensure the position of UAV to be approximately static over each time slot. 
Moreover, the UAV is assumed to fly at a fixed altitude of $z$ in meters, which stands for the minimum altitude for avoiding collision with ground obstacles. Subsequently, a two dimensional Cartesian coordinate system is considered to measure the horizontal coordinate of all nodes. Let $\mb R = \{\mb r[n]| \forall n \in \mc N\}$ be the UAV's trajectory, where $\mb r[n] \in \mathbb{R}^2$ denotes the horizontal coordinate of the UAV at slot $n$ measured in units of meters. The initial position of UAV is denoted as $\mb r_0$. Regarding wireless links between the UAV and LUs (or Eve) are assumed to be dominated by LoS; and the Doppler impacts caused by UAV movement are able to be well compensated \cite{hua2019energy}. 
Thus, the CSI between the UAV and LU $k$ at slot $n$ is modeled as 
 $
   h_k[n] = \frac{h_0}{(d_k[n])^2} = \frac{h_0}{z^2 + \|\mb r[n] - \mb r_k\|^2} , 
 $
 where $d_k[n]$ denotes the distance between the UAV and LU $k$; $\mb r_k$ denotes the coordinate of LU $k$; and $h_0$ stands for reference channel gain at 1 m \cite{hua2019energy,lee2018uav}. Similarly, the CSI between the UAV and Eve at slot $n$ is given by
$
       h_e[n] = \frac{h_0}{z^2 + \|\mb r[n] - \mb r_e\|^2}, 
$
 where $\mb r_e$ denotes the coordinates of Eve \cite{hua2019energy,lee2018uav}. 


Let $\mb E = \big[e_k[n]\big] \in \{0,1\}^{K \times N}$ be user association strategy. Specifically, if LU $k$ is served by the UAV at slot $n$, we have $e_k[n] = 1$, otherwise, $e_k[n] = 0$. To avoid interference when serving multiple contents, the UAV is allowed to communicate at most one LU at each slot, i.e.,  
$
  \sum_{k \in \mc K}  e_k[n] \leq 1, ~\forall n.
$ 
Consequently, the achievable data rate for LU $k \in \mc K$ at slot $n$ in bps is given by
\begin{align} 
R_k[n] = e_k[n]\log_2 \left(1 + \frac{Ph_0/\sigma^2}{\left(z^2+\|\mb r[n] - \mb r_k\|^2\right)}\right), \forall n, k,
\end{align}  
where $P$ denotes the transmission power; and $\sigma^2$ denotes the variance of additive Gaussian noise. 
Similarly, the achievable data rate at slot $n$ for Eve wiretapping LU $k$ is given by
\begin{align} 
R_{e,k}[n] = e_k[n] \log_2 \left(1 + \frac{Ph_0/\sigma^2}{\left(z^2+\|\mb r[n] - \mb r_e\|^2\right)}\right), \forall n,k. \label{eq:rek}
\end{align}
Accordingly, on the basis of PLS \cite{wyner1975wire}, we apply the Wyner's wiretap code and introduce redundant information to confound Eve. Thus, the achievable average secrecy data rate for LU $k \in \mc K$ at slot $n$ in the presence of Eve is given by
$
   R_k^{\rm sec} [n] = \left[ R_k [n] - R_{e,k}[n]\right]^+, \forall n,k,
$
where $[\cdot]^+$ represents $\max \{\cdot, 0\}$.



\section{Problem Formulation \& Proposed Design}
Our objective is to minimize transmission delay (e.g., $T = N\tau$) for all LUs while guaranteeing secure delivery. Thus, it is necessary to jointly optimize UAV's trajectory $\mb r$, and user association $\mb E$, which leads to the following optimization problem:
\begin{subequations}
  \label{pro:p}
  \begin{align} 
  \mc P: ~~\min_{\mb R, \mb E, N}~ &N \tau\\
  s.t. ~  
  & B_0\tau\textstyle\sum_{n \in \mc N} R^{\rm sec}_k[n] \geq s, \forall k, \label{eq:rate}\\
  & \|\mb r[n+1] - \mb r[n]\|_2 \leq v_{\max} \tau, \forall n, \label{eq:r1}\\
  & \mb r[1] = \mb r_0, \mb r[N] = \mb r_0, \label{eq:r2}\\
  & \textstyle\sum_{k \in \mc K} e_{k} [n]  \leq 1, ~\forall n, \label{eq:ts1}\\
  & e_k[n] \in \{0,1\}, \forall k, \forall n, \label{eq:ts2}
  \end{align}
\end{subequations}  
where constraint \eqref{eq:rate} indicates the completion of content delivery, and $B_0$ is the system bandwidth; and constraints \eqref{eq:r1} and \eqref{eq:r2} stand for practical restrictions on UAV movements, with $v_{\max}$ being the maximum speed of the UAV. 

Although problem $\mc P$ is MINLP, we present the characterization for an optimal solution $N^*$ to problem $\mc P$.
\begin{Prop}
   There exists a unique optimal solution $N^*$ satisfying 
    \begin{align}
  N^* \leq \left\lceil 
  \frac{sK} {B_0\tau  \log_2 \frac{(z^2 + \rho_0)(z^2 + \delta_e)}
  {z^2( z^2 + \rho_0 + \delta_e )}}
  + \frac{\|\mb r_K -\mb r_0\| + \sum_{k\in \mc K} \delta_k}{v_{\max}\tau}
  \right\rceil, 
  \label{eq:maxN}
  \end{align}
where $\rho_0 = \frac{Ph_0}{\sigma^2}$; $\delta_e = \max_k \|\mb r_e - \mb r_k\|^2$; $\delta_k = \|\mb r_k - \mb r_{k-1}\|, \forall k \in \mc K$; and $\left\lceil \cdot \right\rceil$ denotes the ceil operation; 
and the right-hand side of \eqref{eq:maxN} is also a feasible solution to $\mc P$.
\end{Prop}
{\it Proof:} The proof of Proposition 1 is omitted due to the page limit. 

It can be observed that the dimension of variables in problem $\mc P$ (related to $N$) is not deterministic. To handle this difficulty, we resort to bisection approach. By fixing $N$, problem $\mc P$ reduces to find a feasible point to constraints \eqref{eq:rate}--\eqref{eq:ts2}. Subsequently, we consider the following problem:
\begin{subequations}
  \label{pro:p}
  \begin{align} 
  \mc P(N): ~~\max_{\mb R, \mb E, \lambda }~ &\lambda \\
  s.t. ~ 
  & \textstyle\sum_{n \in \mc N} R^{\rm sec}_k[n] \geq \lambda, \forall k \in \mc K, \label{eq:rnc}\\
  & \eqref{eq:r1} - \eqref{eq:ts2}. 
  \end{align}
\end{subequations} 
Denote the optimal value for problem $\mc P(N)$ as $\lambda^*$. It is straightforward to see that if $B_0\tau\lambda^* \geq s$, an optimal solution to $\mc P(N)$ is a feasible solution to the original $\mc P$. Note that, to deal with secrecy rate function in constraint \eqref{eq:rnc}, we consider the following problem:
\begin{subequations}
  \label{pro:pp}
  \begin{align} 
  \max_{\mb R, \mb E, \lambda }~ &\lambda \\
  s.t. ~ 
  & \sum_{n =1}^{N} (R_k[n] - R_{e,k}[n]) \geq \lambda, \forall k \in \mc K, \label{eq:rnc2}\\
  & \eqref{eq:r1} - \eqref{eq:ts2}. 
  \end{align}
\end{subequations} 

\begin{Prop}
  Problem \eqref{pro:pp} attains the same optimal value as that of problem $\mc P {(N)}$ . 
\end{Prop}
{\it Proof:} By applying the idea of the proof for Lemma 1 in \cite{zhang2019securing}, we can always set $e_k[n]$ to be zero if $R_k[n] - R_{e,k}[n]$ is negative. 

Now, we turn our attention to tackle the binary variables in problem \eqref{pro:pp}. Instead of simply relaxing $\mb E$ into a continuous one, a variational approach is introduced.  
\begin{Lemma} {\rm (\cite{wu2019jointMDS})}
  Define set $\Omega = \{ (\mb E, \mb X)| {\rm tr}\{(2\mb E - \mb 1)(2\mb X - \mb 1)^T\} = KN, \mb 0 \leq \mb E \leq \mb 1, \|2\mb X - \mb 1\|_F^2 \leq KN, \mb X \in \mathbb R^{K\times N}\}$. Then, for any $(\mb E, \mb X) \in \Omega$, it satisfies $\mb E \in \{0,1\}^{K \times N}$ and $\mb X = \mb E$. 
\end{Lemma}
By leveraging Lemma 1, problem \eqref{pro:pp} can be  reformulated as
  
  \begin{align} 
  \label{pro:ppp}
  \max_{\mb R, \mb E, \mb X, \lambda }~ \lambda ~~
  s.t. ~ (\mb E, \mb X) \in \Omega, \eqref{eq:r1} - \eqref{eq:ts1}, \eqref{eq:rnc2}.
  \end{align}
Although now problem \eqref{pro:ppp} becomes continuous, $\Omega$ is still nonconvex. Thus, we consider constructing the following penalty function:
\begin{align} 
  h(\mb E, \mb X) = KN - {\rm tr}\{(2\mb E - \mb 1)(2\mb X - \mb 1)^T\},  \label{eq:pen}
\end{align}
which is bilinear \cite{yuan2016binary}. 
Thus, by introducing a penalty parameter $\omega >0$, problem \eqref{pro:ppp} can be addressed by solving the following problem:
\begin{subequations}
  \begin{align} 
  \mc P_1(N): ~~\max_{\mb R, \mb E, \mb X, \lambda }~ &\lambda - \omega h(\mb E, \mb X)\\
  s.t. ~ 
  & \mb 0 \leq \mb E \leq \mb 1, \|2\mb X - \mb 1\|_F^2 \leq KN, \label{eq:ex2}\\
  &  \eqref{eq:r1} - \eqref{eq:ts1}, \eqref{eq:rnc2}.
  \end{align}
\end{subequations} 
It can be shown that $h(\mb E, \mb X) \geq 0$ holds for any point within constraint \eqref{eq:ex2}. Moreover, a  feasible point $(\mb E^*, \mb X^*)$ to problem \eqref{pro:ppp} always satisfies $h(\mb E^*, \mb X^*) = 0$; and other points $(\mb E, \mb X)$ feasible to constraint \eqref{eq:ex2} may give rise to $h(\mb E, \mb X) > 0$. Hence, the violation of penalty function can always be penalized by using a proper $\omega$ so as to generate a binary solution for user association.  

To take advantage of the nice structure of the penalty function, $\mc P_1(N)$ can be first decomposed into two subproblems. \\
\underline{\it Trajectory Design with Fixed User Association:} 
The subproblem for trajectory design with fixed $(\mb E, N)$ is given by
\begin{subequations}
  \label{pro:sub1}
  \begin{align} 
 \max_{\mb R, \mb X, \lambda }~ &\lambda - \omega h(\mb E, \mb X)\\
  s.t. ~ 
  & \|2\mb X - \mb 1\|_F^2 \leq KN, \label{eq:ex3}\\
  &  \eqref{eq:r1}, \eqref{eq:r2},\eqref{eq:rnc2}.
  \end{align}
\end{subequations} 
It can be seen that nonconvex constraint \eqref{eq:rnc2} constitutes a hinder to address problem \eqref{pro:sub1}. Note that, $R_{e,k}[n]$ (e.g., defined as \eqref{eq:rek}) can be re-expressed as 
\begin{align} 
  R_{e,k}[n] = e_k[n] \log_2\left( 1 + \frac{\rho_0}{z^2 + d_e[n]}\right), 
\end{align}  
where the snack variables $\mb d_e = \{d_e[n]\}$ satisfy
$
   d_e[n] = \|\mb r[n] - \mb r_e\|^2. 
$
Clearly, $R_{e,k}[n]$ is convex w.r.t. $d_e[n]$. 
Without loss of optimality, subproblem \eqref{pro:sub1} can be relaxed as
\begin{subequations}
  \begin{align} 
  \mc R_1(\mb E, N): ~~\max_{\mb R, \mb X, \mb d_e, \lambda }~ &\lambda - \omega h(\mb E, \mb X)\\
  s.t. ~ 
  & d_e[n] \leq \|\mb r[n] - \mb r_e\|^2, \forall n, \label{eq:den}\\
  & \eqref{eq:r1}, \eqref{eq:r2}, \eqref{eq:rnc2}, \eqref{eq:ex3},  
  \end{align}
\end{subequations}
where constraint \eqref{eq:rnc2} and \eqref{eq:den} are nonconvex. 
Hereunder, we claim the characterization for an optimal solution to problem $\mc R_1(\mb E, N)$.
\begin{Lemma}
  {\rm (\cite{wu2019jointMDS})} An optimal solution $\mb X^*$ to $\mc R_1(\mb E, N)$ is given by
  \begin{align} 
    \mb X^* = \begin{cases}
    {any~feasible~value}, & if~ \mb E = {\bf \frac{1}{2}},\\
    \frac{\sqrt{KN}(2 \mb E -1)}{2 \|2\mb E -\mb 1\|_F} + {\bf \frac{1}{2}}. 
    \end{cases} 
    \label{eq:X}
  \end{align}  
\end{Lemma}
\begin{Prop}
  Define the following function
  $
    f(\mb y) = \log_2( 1 + \frac{a_1}{a_2 + \|\mb y\|^2}), \forall \mb y,
 $
  where coefficients $a_1, a_2 > 0$. Then, it holds that 
  \begin{align*} 
    f(\mb y) \geq  f(\mb y_0) - \frac{a_1(\|\mb y\|^2 - \|\mb y_0\|^2)}{(a_1 + a_2 + \|\mb y_0\|^2)(a_2+ \|\mb y_0\|^2) \log2} \triangleq \underline f(\mb y), 
  \end{align*}  
for any point $\mb y_0$, where $\underline f(\mb y)$ is concave, and the equality meets when $\mb y = \mb y_0$.  
\end{Prop}
Accordingly, we have $R_k[n] = e_k[n]f(\mb r[n] -\mb r_k|a_{1}, a_{2})$, where coefficients $a_{1} = \rho$ and $a_{2} = z^2$. By leveraging Proposition 3, it results in  $R_k[n] \geq \underline R_k[\mb r_0, n] \triangleq e_k[n] \underline f(\mb r[n] -\mb r_k|a_{1}, a_{2}, \mb r_{0}[n])$
 where $\mb r_{0}[n]$ is any local point, and thereby the equality holds if $\mb r[n] = \mb r_{0}[n]$. To make constraint \eqref{eq:den} tractable, we consider using the first order Taylor expansion to lower bound the right-hand side of \eqref{eq:den}. Thus, a minorant function to $\|\mb r[n] - \mb r_e\|^2$ is given by
$
    \underline d_e[\mb r_0, n] \triangleq 2 (\mb r[n] - \mb r_0[n])^T (\mb r_0[n] - \mb r_e) + \|\mb r_0[n] - \mb r_e\|^2.
 $ Thus, $ \|\mb r[n] - \mb r_e\|^2 \geq \underline d_e[\mb r_0, n]$, and the equality meets again if $\mb r[n] = \mb r_{0}[n]$. As a result, problem $\mc R_1(\mb E, N)$ can be solved by addressing its inner approximation problem: 
  \begin{align*} 
  \mc R_2(\mb E, N, \mb r_0): \max_{\mb R, \mb d_e, \lambda }~ &\lambda\\
  s.t. ~ 
  & d_e[n] \leq \underline d_e[\mb r_0, n], \forall n,\\
  & \textstyle\sum_{n =1}^{N} (\underline R_k[\mb r_0, n] - R_{e,k}[n]) \geq \lambda, \forall k, \\
  & \eqref{eq:r1}, \eqref{eq:r2}. 
  \end{align*}
which is convex and can be solved by interior-point method.  \\
\underline{\it User Association Design with Fixed Trajectory:} 
The subproblem for user association design with fixed $(\mb R, \mb X)$ is given by 
  \begin{align*} 
  \mc R_3(\mb R, \mb X, N): ~\mathop{\max}\limits_{\mb E, \lambda}~ &\lambda - \omega h(\mb E, \mb X)~s.t. ~\mb 0 \leq \mb E \leq \mb 1, \eqref{eq:ts1}, \eqref{eq:rnc2},
  \end{align*}
which is a linear programming and can be easily solved by applying interior-point method. \\
\underline{\it Proposed Algorithms:} The proposed algorithm to solve subproblem $\mc P_1(N)$ is presented in Algorithm 1. Notably, Algorithm 1 operates in an inexact block coordinate descent (BCD) fashion, i.e., the block variable $(\mb R, \mb E)$ is not updated optimally in each iteration. Moreover, the auxiliary variable $\mb X$ is updated with closed-form expression. These features help to reduce computational complexity. 
Then, the proposed design for transmission latency minimization problem $\mc P$ is shown as Algorithm 2. To reduce iterations, a feasible $N_{\max}$ should be selected according to Proposition 1.  
\begin{algorithm}[h]
\caption{Inexact BCD Penalty Design for Problem $\mc P_1(N)$}
\begin{algorithmic}
\State {\bf Initialize} $ i = 0, N, \omega>0, \mb E^0,\mb R^0$
\Repeat
\State solve $\mc R_2(\mb E^{(i)}, N, \mb R^{(i)})$ to attain $\mb R^{(i+1)}$
\State attain $\mb X^{(i+1)}$ according to \eqref{eq:X} 
\State solve $\mc R_3(\mb R^{(i+1)}, \mb X^{(i+1)}, N)$ to attain $\mb E^{(i+1)}$
\State update $i \leftarrow i+1$
\Until{some stopping criterion is met}
\end{algorithmic}
\end{algorithm}  
\begin{algorithm}[h]
\caption{Bisection Algorithm for Problem $\mc P$}
\begin{algorithmic}[1]\label{Ag1}
\State {\bf Initialize} $N_{\min}, N_{\max}$
\Repeat
\State $N = \lceil(N_{\min} + N_{\max})/2\rceil$
\State solve problem $\mc P(N)$ and obtain $\lambda^*$
\If{$\lambda^* \tau B_0 \geq s$}
\State $N_{\max} = N$
\Else
 \State $N_{\min} = N$
\EndIf
\Until{$N_{\max} - N_{\min} =1 $}
\end{algorithmic}
\end{algorithm}

\section{Performance Evaluation}
In this section, we present numerical simulations to evaluate the performance of the proposed scheme. We consider the following setting: the UAV is initially located at $\mb r_0 = (400~{\rm m}, 200~{\rm m})$, and flies at a fixed altitude $z = 100~{\rm m}$; the maximum flying speed of the UAV is $50 ~{\rm m/s}$ \cite{hua2019energy}; transmission power $P = 0$ dB; reference channel gain and noise power in turn are $-60$ dB and $-110$ dBm; 2 LUs are considered and distributed at $\mb r1 = (200~{\rm m}, 0~{\rm m})$ and $\mb r2 = (600~{\rm m}, 0~{\rm m})$, respectively; each user requests certain content item of equal size in $10$ MB; Eve's position is $(500~{\rm m}, 100~{\rm m})$; and the during of each discrete time is $0.5$ s \cite{hua2019energy}.  

\begin{figure}[h]
  \centering
  \includegraphics[scale=0.28]{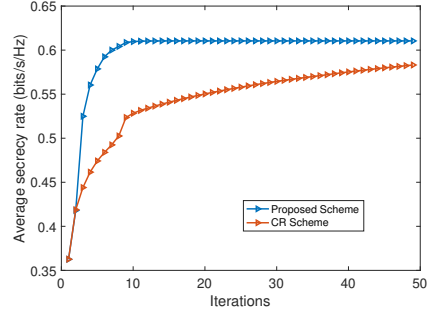}
  \caption{Convergence behavior of Algorithm 1.}
  \label{fig:conv}
\end{figure}

\begin{figure}[h]
  \centering
  \includegraphics[scale=0.28]{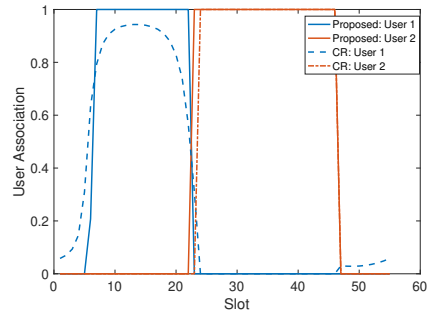}
  \caption{User association under different schemes.}
  \label{fig:us}
\end{figure}

To illustrate the superior advantage of the proposed variational approach (as Algorithm 1) over the continuous relaxation (CR) scheme (relaxing binary variables into continous ones; see \cite{wu2018joint}), we plot the convergence behavior in Fig. \ref{fig:conv}. A preset number of transmission slots is given by $N = 55$ according to Proposition 1; we empirically set the penalty parameter $\omega = 0.1$ to prevent the penalty term from dominating the total objective value. 
It can be seen that the proposed scheme converges faster than the CR scheme and reaches the maximum secrecy rate within 10 iterations. 
Moreover, after it reaches the maximum iteration, the proposed scheme can obtain 4.7\% higher secrecy rate than that of the CR scheme.
To further illuminate the potentials of the proposed approach, we depict user association strategies for both schemes in Fig. \ref{fig:us}. Clearly, the proposed scheme can result in almost binary solutions for each user while CR witnesses a large number of {\it fractional} decision variables. These fractional solution needs to be further dealt with by partitioning each discrete slot into multiple sub-slots \cite{wu2018joint}. By using the proposed scheme, content transmissions are able to be completed at $N = 46$, resulting in a lower transmission delay. Obviously, the expansiveness of scheduling time for user 1 resulting from the CR scheme leads to a longer content download. The above comparisons demonstrate that the proposed scheme exhibits superior performance over the extant scheme. 
\begin{figure}[!h]
  \centering
  \includegraphics[scale=0.28]{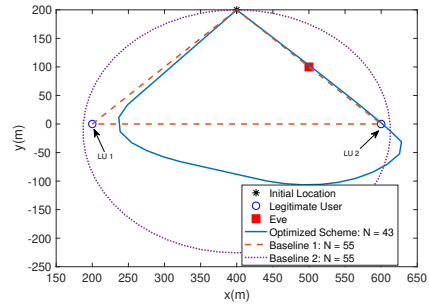}
  \caption{UAV trajectory under different schemes.}
  \label{fig:tra}
\end{figure}

The optimized trajectory is depicted in Fig. \ref{fig:tra}, which is obtained by using Algorithm 2. In particular, baseline 1 presents the result for traditional hover scheme: the UAV first finds the shortest path to the LU, then hovers exactly above the LU to serve the desired content, and flies to next LU after finishing content delivery. Moreover, baseline 2 is realized by following circular trajectory where the number of transmission slots is the initial point selected by Proposition 1. Obviously, without optimization, both baselines exhibit long delay, i.e., $N = 55$. Regarding the optimized scheme, it finishes in $N = 43$. Specifically, the UAV first flies close to LU 1, and then deviates from the straight line to approach LU 2 so as to provide high secrecy data rate; when the UAV finishes content transmission, it directly flies back to the starting point.

\section{Conclusion}
In this work, the security of UAV-aided communications has been investigated with regard to minimizing transmission latency. We have studied a scenario where the UAV is utilized to satisfy users' content requests in the presence of an eavesdropper. The formulated problem is a mixed integer nonlinear program. 
We have developed a bisection approach to efficiently address this problem. In particular, the associated subproblem has been addressed by leveraging a variational penalty method. Simulation results have indicated that the proposed design under the scenario being studied can achieve 4.7\% higher average secrecy rate than the continuous relaxation approach widely adopted in prior works, leading to a low-latency transmission scheme. 
\bibliographystyle{IEEEtran}
\bibliography{references}
\end{document}